\documentclass[%
reprint,
amsmath,amssymb,
aps,
]{revtex4-2}

\usepackage{graphicx}% Include figure files
\usepackage{dcolumn}% Align table columns on decimal point
\usepackage{bm}% bold math
\usepackage{siunitx}% use SI units
\usepackage{hyperref}% add hypertext capabilities
\usepackage{comment}
\usepackage{color}
\usepackage{graphicx}
\usepackage{mathpazo}
\usepackage{ulem}
\usepackage{overpic}
\usepackage{float}
\usepackage{placeins}

\begin{document}
	
	\preprint{APS/123-QED}
	
	\title{Temperature dependence of charge conversion during NV-center relaxometry}% Force line breaks with \\
	\author{Isabel Cardoso Barbosa}
	\affiliation{
		Department of Physics and State Research Center OPTIMAS, University of Kaiserslautern-Landau, Erwin-Schroedinger-Str. 46, 67663 Kaiserslautern, Germany\\
	}

	\author{Jonas Gutsche}
	\affiliation{
		Department of Physics and State Research Center OPTIMAS, University of Kaiserslautern-Landau, Erwin-Schroedinger-Str. 46,
        67663 Kaiserslautern, Germany\\
	}
	
	\author{Dennis L\"onard}
	\affiliation{
		Department of Physics and State Research Center OPTIMAS, University of Kaiserslautern-Landau, Erwin-Schroedinger-Str. 46,
        67663 Kaiserslautern, Germany\\
	}
	
	\author{Stefan Dix}
	\affiliation{
		Department of Physics and State Research Center OPTIMAS, University of Kaiserslautern-Landau, Erwin-Schroedinger-Str. 46,
        67663 Kaiserslautern, Germany\\
	}
	
	\author{Artur Widera}
	\email{Author to whom correspondence should be addressed: widera@physik.uni-kl.de}
    \affiliation{
		Department of Physics and State Research Center OPTIMAS, University of Kaiserslautern-Landau, Erwin-Schroedinger-Str. 46,
        67663 Kaiserslautern, Germany\\
	}
	
	\date{\today}% It is always \today, today,
	
	\begin{abstract}
    Temperature-dependent nitrogen-vacancy (NV)-center relaxometry is an established tool to characterize paramagnetic molecules near to a sensing diamond, gaining momentum in different fields of science. 
    However, recent results indicate that conversion between NV-center charge states impedes these measurements and influences the results for the $T_1$ time. 
    While the temperature dependence of NV centers' $T_1$ time is well-studied, additional contributions from temperature-dependent charge conversion during the dark time may further affect the measurement results. 
    We combine temperature-dependent relaxometry and fluorescence spectroscopy at varying laser powers to unravel the temperature dependence of charge conversion in nanodiamond for biologically relevant temperatures. 
    While we observe a decrease of the $T_1$ time with increasing temperatures, charge conversion remains unaffected by the temperature change. 
    These results allow the temperature-dependent performance of $T_1$ relaxometry without further consideration of temperature dependence of charge conversion. 
	\end{abstract}
	
	\maketitle
	
	%\tableofcontents
	
	\section{\label{sec:level1}INTRODUCTION}
	
    The negatively charged NV center in diamond is an established tool for spatially-resolved mapping of temperatures \cite{Kucsko.2013}, magnetic fields \cite{Rondin.2014}, and electric fields \cite{Dolde.2011} as so-called quantum sensors \cite{Barry.2020}. 
    Because of their small sizes, nanodiamonds provide an advantageous NV host for different fields of science.
    Nanodiamonds can be brought into cells or organelles \cite{Sharmin.2021, Sigaeva.2022, Mzyk.2022} and trace chemical reactions via NV-center relaxometry \cite{Barton.2020, Nie.2021}, which is the measurement of the NV centers' longitudinal spin-relaxation time $T_1$ \cite{Levine.2019}. 
    In a chemical reaction, not only the formation of a paramagnetic product, but also changes in the temperature or the pH may contribute to a $T_1$ change \cite{Nie.2021}.  
    Moreover, recent studies indicate that another mechanism challenges $T_1$ relaxometry: the illumination-induced charge conversion between NV-charge states during a pulsed-laser measurement \cite{Giri.2018, Giri.2019, CardosoBarbosa.2023}. 
    These ionization processes between the negatively charged $\mathrm{NV}^-$ and the neutrally charged $\mathrm{NV}^0$ impact the $\mathrm{NV}^-$ fluorescence recorded during the $T_1$ measurement.
    With temperatures potentially increasing during reactions, the question arises whether an additional contribution due to temperature-dependent charge conversion must be considered when interpreting changes in the relaxometry signal.
    
    In our work, we perform NV fluorescence spectroscopy to derive ratios of $\mathrm{NV}^-$ and $\mathrm{NV}^0$ and compare them for different temperatures. 
    We perform relaxometry at different laser powers to derive the longitudinal spin-relaxation times for an NV ensemble in nanodiamond and the recharge times from $\mathrm{NV}^0$ to $\mathrm{NV}^-$ in the dark for biologically-relevant temperatures \cite{Chretien.2018, Nie.2021} from \SI{20}{\degreeCelsius} to \SI{75}{\degreeCelsius} (\SI{294}{\kelvin} to \SI{348}{\kelvin}).
    The combination of these two techniques allows us to trace the ratio $[\mathrm{NV}^-]/[\mathrm{NV}^0]$ during the relaxometry sequence as a function of the temperature.
    
    We non-resonantly excite the negatively charged NV center ensemble in our sample with a \SI{520}{\nano\meter} laser from their triplet ground state to their triplet excited state \cite{Doherty.2011, Barry.2020}. 
    Due to the spin-state dependent transition rates between the $\mathrm{NV}^-$ states \cite{Robledo.2011, Tetienne.2012}, spin polarization toward the $m_S = 0$ state is achieved \cite{Levine.2019}. 
    This spin polarization relaxes over the typical time scale $T_1$ until the population decays to a thermally mixed state \cite{Levine.2019}.
    During laser illumination, ionization from $\mathrm{NV}^-$ to $\mathrm{NV}^0$ can take place \cite{Manson.2005, Aslam.2013, Manson.2018, Dhomkar.2018}, and recharging from $\mathrm{NV}^0$ to $\mathrm{NV}^-$ in the dark then contributes to the measured $\mathrm{NV}^-$ fluorescence signal \cite{Giri.2018, Giri.2019, CardosoBarbosa.2023}.
    While common $T_1$ measurement schemes include a microwave-$\pi$ pulse between laser pulses to mitigate influences from background fluorescence \cite{Jarmola.2012, Mrozek.2015}, the microwave pulse is often omitted in biological applications to avoid undesired heating of samples \cite{Barton.2020, Sigaeva.2022, Mzyk.2022}.
    Therefore, these all-optical $T_1$ measurements are susceptible to influences from charge conversion.
    With the temperatures potentially rising during chemical reactions, the $\mathrm{NV}^-$ fluorescence signal could be influenced by $T_1$'s temperature dependence \cite{Jarmola.2012, Mrozek.2015, Norambuena.2018} and charge conversion changing with increasing temperature.
    
	\section{\label{sec:level2}METHODS}	
    We excite ensembles of NV centers in a single nanodiamond of size $\sim \SI{750}{\nano\meter}$ (Adamas Nano, NDNV/NVN700nm2mg) with a \SI{520}{\nano\meter} laser and collect their fluorescence.
    According to the manufacturer, the nanodiamonds contain $\sim$~\SI{0.5}{ppm} of NV centers and $\sim$~\SI{37}{ppm} of substitutional nitrogen. 
    The laser beam is focused to a spot size of $\sim$ \SI{700}{\nano\meter} ($1/e^2$ diameter) with a maximum laser power of \SI{4}{\milli\watt}, and we adjust lower laser powers with an acousto-optic modulator (AOM).
    We simultaneously record fluorescence spectra of $\mathrm{NV}^-$ and $\mathrm{NV}^0$ in the range between \SI{550}{\nano\meter} and \SI{775}{\nano\meter}. 
    Moreover, we decompose and analyze the fluorescence spectra according to Ref. \cite{Alsid.2019}, allowing us to calculate fractions of $[\mathrm{NV}^-]$ and ratios of $[\mathrm{NV}^-]/[\mathrm{NV}^0]$ at specific laser powers and temperatures.
    Additional details on the spectral analysis in this work are given in Appendix~\ref{subsec:decomposition}.
    We perform a common relaxometry sequence for NV-center relaxometry, including \SI{520}{\nano\meter}-laser pulses generated with an AOM and a microwave-$\pi$ pulse. 
    
    To detect the fluorescence of $\mathrm{NV}^-$ and $\mathrm{NV}^0$ separately in two single-photon-counting modules (SPCMs), we split the fluorescence and equip the beam paths with a \SI{665}{\nano\meter} longpass filter ($\mathrm{NV}^-$ fluorescence) and a \SI{600}{\nano\meter} shortpass filter ($\mathrm{NV}^0$ fluorescence). 
    Details on the experimental setup and the pulsed sequence are described in \cite{CardosoBarbosa.2023}.
    Using the described method in \cite{CardosoBarbosa.2023}, we map the NV-charge-state ratio obtained from fluorescence spectroscopy to count-rate ratios in the SPCMs. 
    This mapping allows us to trace the ratio $[\mathrm{NV}^-]/[\mathrm{NV}^0]$ during the relaxometry sequence and get an insight into charge conversion within the relaxation time.
    In this analysis, we correct fluorescence counts for the wavelength-dependent absorption of neutral density filters we employ to keep the SPCMs below saturation.
    Additionally, we consider dark counts recorded with each SPCM.
    Microwave signals are generated, amplified, and brought close to the nanodiamond with a microwave antenna which is similar to the design described in Ref. \cite{Opaluch.2021}.
    We employ a temperature-controlled sample holder to perform temperature-dependent fluorescence spectroscopy and relaxometry.
    Built-in Peltier elements are used to heat the sample and keep the temperature constant at temperatures between room temperature and $\sim \SI{350}{\kelvin}$.
    We derive the temperature at the nanodiamond location via optically-detected magnetic resonance (ODMR) spectroscopy employing the temperature dependence of the NVs' zero-field splitting (ZFS) of \SI{-74.2 \pm 0.7}{\kilo\hertz\per\kelvin} \cite{Acosta.2010}.
    We perform all experiments in a magnetic bias field in the order of \SI{8}{\milli\tesla} to split the NV centers' spin resonances and avoid cross relaxations \cite{Jarmola.2012}.

	\section{\label{sec:Results}RESULTS AND DISCUSSION}
	\subsection{\label{subsec:Spec_analysis}Temperature-dependent spectra decomposition}
	\begin{figure}[b]
            \includegraphics[width=86mm]{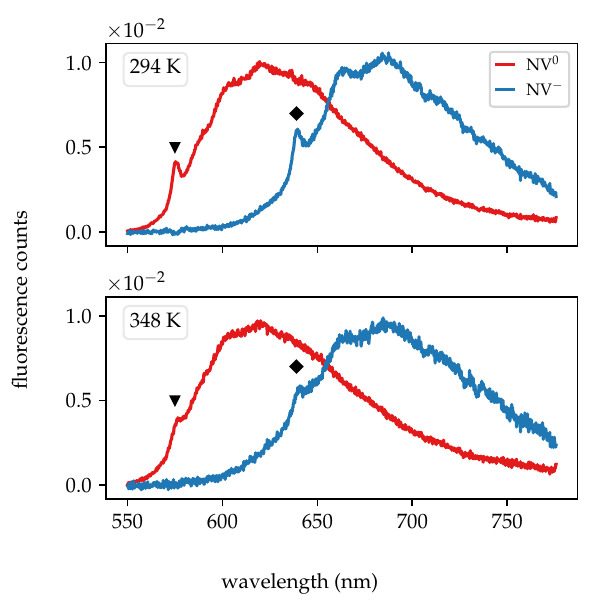}
            \caption{\label{fig:spectra_T}
            Decomposed fluorescence spectra for $\mathrm{NV}^-$ and $\mathrm{NV}^0$ for temperatures of \SI{294}{\kelvin} and \SI{348}{\kelvin}. The ZPLs of $\mathrm{NV}^-$ and $\mathrm{NV}^0$ at \SI{575}{\nano\meter} ($\blacktriangledown$) and \SI{639}{\nano\meter} ($\blacklozenge$), respectively, are broadened and shifted to higher wavelengths at \SI{348}{\kelvin} compared to \SI{294}{\kelvin}.
            }
    \end{figure}
    The decomposed fluorescence spectra of $\mathrm{NV}^-$ and $\mathrm{NV}^0$ are shown in Fig.~\ref{fig:spectra_T} for temperatures of \SI{294}{\kelvin} and \SI{348}{\kelvin}.
    We observe an increase in the $\mathrm{NV}^-$ and $\mathrm{NV}^0$ zero-phonon line (ZPL) widths and a shift of their positions to higher wavelengths caused by the temperature increase as previously described by \cite{Fu.2009, Chen.2011, Doherty.2014, Hui.2019, Yang.2023}, see also Appendix~\ref{subsec:ZPL}.
    The fractional concentration of $\mathrm{NV}^-$ with respect to the sum of $[\mathrm{NV}^-]$ and $[\mathrm{NV}^0]$ is derived by the fractional contributions $c_-$ and $c_0$ of $\mathrm{NV}^-$ and $\mathrm{NV}^0$ to the composed fluorescence spectrum with
    \begin{equation}
        \frac{[\mathrm{NV}^-]}{[\mathrm{NV}^-]+[\mathrm{NV}^0]} = 
        \frac{c_-}{c_- + \kappa_{520} c_0},
        \label{eq:frac}
    \end{equation}
    where $c_- + c_0 = 1$, and $\kappa_{\lambda}$ (here, $\lambda = \SI{520}{\nano\meter}$) is a correction factor that describes the photoluminescence ratios of $\mathrm{NV}^-$ and $\mathrm{NV}^0$ \cite{Alsid.2019}.
    In the spectral analysis, we calculate $\kappa_{520}$ similarly as described in Ref.~\cite{Alsid.2019} and our previous work \cite{CardosoBarbosa.2023}. 
    For this, we record fluorescence spectra at laser powers well below saturation \cite{Wolf.2015} and correct them for the exposure times they were recorded with. 
    We decompose the spectra into contributions from $\mathrm{NV}^-$ and $\mathrm{NV}^0$ and calculate fluorescence intensities for $\mathrm{NV}^-$ and $\mathrm{NV}^0$ for each laser power. 
    We fit the fluorescence intensities of $\mathrm{NV}^-$ and $\mathrm{NV}^0$ as a function of the laser power in a weighted linear fit and directly compare the slopes of the integrated fluorescence intensities of $\mathrm{NV}^-$ and $\mathrm{NV}^0$. 
    By forming the quotient of these slopes, we obtain $\kappa_{520}$.

    \begin{figure}[]
            \includegraphics[width=86mm]{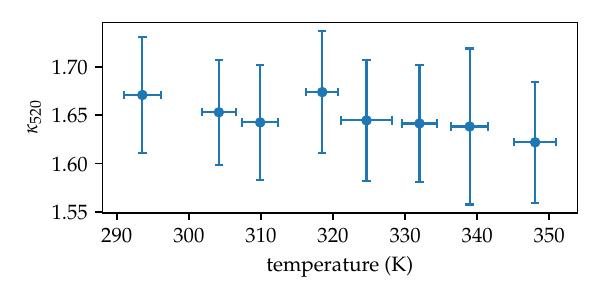}
            \caption{\label{fig:k520_T}
            Calculated $\kappa_{520}$ as a function of the temperature. We derive $\kappa_{520}$ from spectral analysis at each temperature as described in the main text.
            }
    \end{figure}
    
    \begin{figure}[]
            \includegraphics[width=86mm]{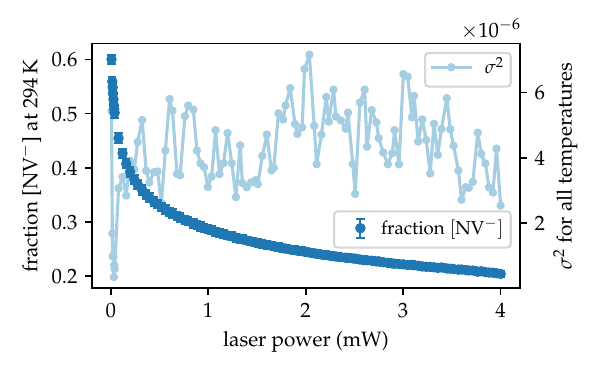}
            \caption{\label{fig:Ratio_T}
            Calculated fraction of $[\mathrm{NV}^-]$ as a function of the laser power at \SI{294}{\kelvin}. Representing increased temperatures up to \SI{348}{\kelvin}, the variance $\sigma^2$ between the $[\mathrm{NV}^-]$ fractions at each temperature is shown as a function of the laser power.
            }
    \end{figure}
    
    We perform this spectral analysis for different temperatures ranging from \SI{294}{\kelvin} to \SI{348}{\kelvin}. 
    The values obtained for $\kappa_{520}$ show a relatively constant value around $\kappa_{520} \sim \SI{1.65}{}$, see Fig.~\ref{fig:k520_T}. 
    The error of the temperatures is derived from the fit error of the resonances measured with ODMR spectroscopy and the uncertainty of the temperature dependence of the NVs' ZFS \cite{Acosta.2010}.
    We derive the standard error for $\kappa_{520}$ from linear fitting of the fluorescence intensities we obtain for $\mathrm{NV}^-$ and $\mathrm{NV}^0$, weighted with the statistical error of the intensities we obtain in the spectral analysis (inverse-variance fit).
    
    Additionally, we record NV fluorescence spectra for laser powers up to \SI{4}{\milli\watt} at each temperature.
    Using the values for $\kappa_{520}$ we obtain for each temperature and the decomposed spectra, we calculate the fractional concentration of $\mathrm{NV}^-$ according to Eq.~(\ref{eq:frac}) for each laser power and temperature.
    We obtain the error for the $[\mathrm{NV}^-]$ fraction from the errors for $\kappa_{520}$, $c_-$, and $c_0$.
    The result for the $[\mathrm{NV}^-]$ fraction as a function of the laser power is shown in Fig.~\ref{fig:Ratio_T} for \SI{294}{\kelvin}.
    At very low laser powers of $\sim$~\SI{8}{\micro\watt}, the fraction of $[\mathrm{NV}^-]$ is above \SI{60}{\percent}, while it decreases to \SI{20}{\percent} at high laser powers of $\sim$~\SI{4}{\milli\watt}. 
    Although the fluorescence spectra in Fig.~\ref{fig:spectra_T} change visibly with increasing temperature, the fractions of $[\mathrm{NV}^-]$ qualitatively show the same behavior as a function of the laser power for all temperatures. 
    To point out differences between temperatures in more detail, we show the variance $\sigma^2$ of the $[\mathrm{NV}^-]$ fractions between all eight temperatures for all laser powers in Fig.~\ref{fig:Ratio_T}.
    With $\sigma^2$ staying in the order of $10^{-6}$ for all laser powers, we confirm that the $\mathrm{NV}^-$ fraction as a function of the laser power does not depend on the temperature in our sample.
	
	\subsection{\label{subsec:Relaxometry}Temperature-dependent relaxometry}
\begin{figure}[]
     \begin{overpic}[width=86mm]{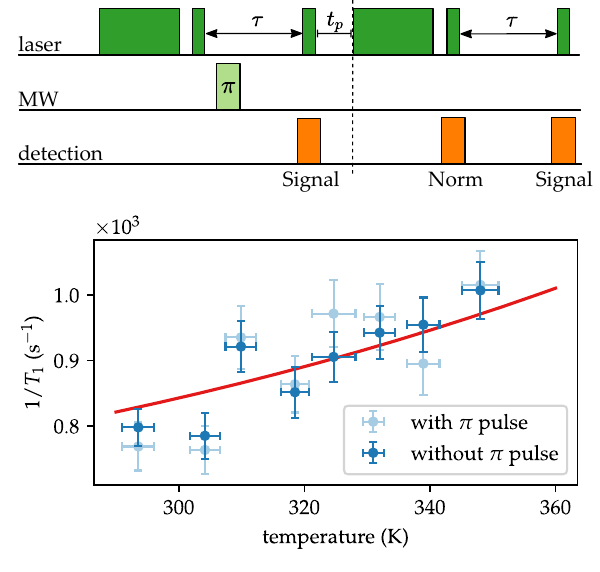}
         \put(0,94){(a)}
         \put(0,56){(b)}
        \end{overpic}
            \caption{\label{fig:T_1_Relaxometry}
            NV-center relaxometry scheme and result for $1/T_1$ as a function of the temperature. 
            (a) The pulsed scheme uses laser spin-polarization and readout pulses. Additionally, we apply a microwave $\pi$ pulse in the first half of the sequence. The $T_1$ time, accessed with the $\pi$ pulse, is obtained by forming the difference of the fluorescence counts in the two signal detection windows and fitting a monoexponential function to the result. Further, with the same measurement, we recover the all-optical $T_1$ time in the second half by dividing the fluorescence counts in the signal detection window by the fluorescence counts in the normalization detection window and fitting a monoexponential function to the result. (b) Result for $1/T_1$ for each temperature, evaluated with the $\pi$ pulse applied and without. The red solid line is a fit as explained in the main text.
            } 
    \end{figure}

	To get an insight into the temperature dependence of the $T_1$ spin-relaxation time and charge conversion in our sample, we perform $T_1$ relaxometry using a common sequence shown in Fig.~\ref{fig:T_1_Relaxometry}~(a) \cite{Jarmola.2012, Mrozek.2015} at different temperatures and laser powers.
	A laser pulse of $\SI{200}{\micro\second}$ spin polarizes the $\mathrm{NV}^-$ centers to their $m_S = 0$ states.
	After \SI{1}{\micro\second}, a $\SI{5}{\micro\second}$ laser pulse probes the fluorescence intensity before the variable relaxation time $\tau$ for normalization.
	The $\pi$ pulse performed during $\tau$ acts upon $\mathrm{NV}^-$ centers of a specific orientation in the diamond lattice and excites their population from $m_S = 0$ to $m_S = +1$ or $m_S = -1$ of the electronic ground state.
	Another $\SI{5}{\micro\second}$ laser pulse reads out the fluorescence intensity after $\tau$ in the signal detection windows.
	Between readout and subsequent spin polarization, a pause time $t_p$ of \SI{1}{\milli\second} is inserted to mitigate build-up effects during the cycle.
	Each cycle is repeated \SI{50000}{} times and the complete sequence is swept multiple times.
	By subtraction of the $\mathrm{NV}^-$ fluorescence in the signal detection windows, we obtain a difference in fluorescence between $m_S = 0$ and $m_S = + 1$ or $m_S = - 1$ states as a measure of the spin polarization of the $\mathrm{NV}^-$ centers excited by the $\pi$ pulse \cite{Jarmola.2012, Mrozek.2015}.
	Using the second half of the sequence, we obtain a normalized fluorescence of $\mathrm{NV}^-$ by division of the signal fluorescence counts by the normalization fluorescence counts for all NV-center orientations. 
	This analysis corresponds to an all-optical $T_1$ measurement scheme as often employed in biology \cite{Nie.2021, Sigaeva.2022, Mzyk.2022}.
    We conduct the sequence with low laser powers of $\sim$ \SI{8}{\micro\watt} to prevent charge conversion from prevailing.
    We fit both decays we obtain from the two evaluation methods, with and without the $\pi$ pulse, with a monoexponential function for each temperature to derive the $T_1$ time for $\mathrm{NV}^-$. 
    The results are shown as $1/T_1$ in Fig.~\ref{fig:T_1_Relaxometry}~(b).
    We derive the error for $1/T_1$ for each temperature from the standard error of $T_1$ we obtain from the exponential fit.
    Within the ranges of their fit errors, $1/T_1$ is equal for both evaluation methods at each temperature. 
    This result concludes that $\mathrm{NV}^-$ spin relaxation is dominant due to the low laser power applied, and charge conversion is negligible.
    Additionally, we observe an increase of $1/T_1$ with the temperature. 
    
    The temperature dependence of $1/T_1$ over a wide temperature range has previously been investigated in Refs. \cite{Jarmola.2012, Mrozek.2015} for NV centers in bulk samples of different defect concentrations and further described in \cite{Norambuena.2018}.
    It was found in Refs. \cite{Jarmola.2012, Mrozek.2015} that the observed behavior can be assigned to two-phonon Raman and Orbach-type processes, and the data fit according to
    \begin{equation}
        \frac{1}{T_1} = A_1(S) + \frac{A_2}{e^{\Delta/kT}-1} + A_3T^5 .
        \label{eq:PRL108}
    \end{equation}
	Here, only the parameter $A_1(S)$ depends on the sample, while $A_2$, $A_3$, and $\Delta$ are universal parameters \cite{Jarmola.2012, Mrozek.2015}.
	Moreover, recent studies link the temperature dependence of the NV centers' photoluminescence intensity and spin contrast to similar electron-phonon interactions \cite{Ernst.2023}.
	We take the values for $A_2$, $A_3$, and $\Delta$ as given in \cite{Jarmola.2012} and fit the function to our measurement data sets. 
	The resulting function with a mean parameter of $A_1$ from the signals with and without the $\pi$ pulse can be seen as a red solid line in Fig.~\ref{fig:T_1_Relaxometry}~(b).
	Qualitatively, this fit models our data well.
	We find $A_1 = \SI{657 \pm 13}{\per\second}$, which generally lies within the orders of magnitude given in Refs. \cite{Jarmola.2012, Mrozek.2015}.
	We note that the value for $A_1$ found for our nanodiamond sample is higher compared to $A_1$ found in a bulk diamond of comparable NV-center concentration \cite{Jarmola.2012, Mrozek.2015}.
	Possible reasons for this discrepancy may be the different alignment of the external magnetic field relative to the NV-center axes in our experiments or a different concentration of substitutional nitrogen atoms in our sample. 
	Additionally, recent studies show that the high-temperature annealing of our sample affects on the NV centers' $T_1$ time \cite{Nunn.2023}, which could also result in a different $A_1$ compared to the samples in Refs. \cite{Jarmola.2012, Mrozek.2015}.

\begin{figure}[]
            \includegraphics[width=86mm]{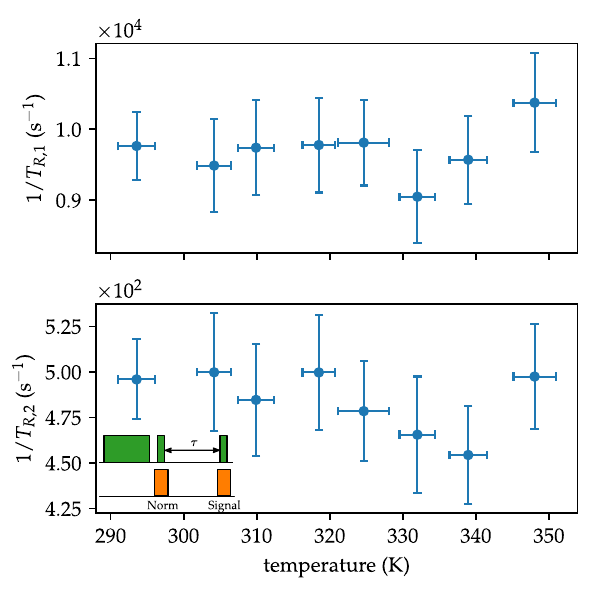}
            \caption{\label{fig:T_R_T}
            Derived values of $1/T_{R,1}$ and $1/T_{R,2}$ as a function of the temperature. We obtain the recharge rates from a biexponential fit to the normalized $\mathrm{NV}^0$ fluorescence, recorded with the relaxometry sequence ($\pi$ pulse omitted) at \SI{0.56}{\milli\watt} laser power. The inset displays the characteristics of the pulsed sequence applied.
            }
    \end{figure}

	To enhance and trace charge conversion during relaxometry, we perform the same sequence in Fig.~~\ref{fig:T_1_Relaxometry}~(a) with a higher laser power of \SI{0.56}{\milli\watt}, omit the microwave $\pi$ pulse, and collect the $\mathrm{NV}^-$ and $\mathrm{NV}^0$ fluorescence in two separate detectors. 
	At high laser powers, $\mathrm{NV}^-$ is converted to $\mathrm{NV}^0$ during the spin polarization pulse \cite{Manson.2005, Giri.2019}.
	During the relaxation time $\tau$, recharge occurs from $\mathrm{NV}^0$ to $\mathrm{NV}^-$, causing the $\mathrm{NV}^0$ fluorescence to decay over $\tau$, while the $\mathrm{NV}^-$ fluorescence increases \cite{Giri.2019, CardosoBarbosa.2023}.
	We interpret this decay of the $\mathrm{NV}^0$ fluorescence in the dark as the recharging time of $\mathrm{NV}^0$ to $\mathrm{NV}^-$ \cite{Giri.2018}. 
	In our nanodiamond sample, the decay of the normalized $\mathrm{NV}^0$ fluorescence cannot be sufficiently described with a monoexponential function \cite{CardosoBarbosa.2023}. 
	Therefore, we fit a biexponential function to the normalized $\mathrm{NV}^0$ fluorescence as a function of $\tau$ to obtain the characteristic recharge times $T_{R,1}$ and $T_{R,2}$ for our sample.
	The result is shown in Fig.~\ref{fig:T_R_T} as $1/T_{R,1}$ and $1/T_{R,2}$.
	We calculate the errors for $1/T_{R,1}$ and $1/T_{R,2}$ from the standard errors we obtain for $T_{R,1}$ and $T_{R,2}$ in the biexponential fit.
    It can be clearly seen in Fig.~\ref{fig:T_R_T} that $1/T_{R,1}$ and $1/T_{R,2}$ are constant as a function of the temperature.
    From this, we conclude that the charge dynamics in our pulsed measurement are independent of the temperature in the observed range.
    In our previous work, several nanodiamonds were investigated with the same measurement scheme and a similar biexponential decay of the $\mathrm{NV}^0$ fluorescence was observed \cite{CardosoBarbosa.2023}. 
    Therefore, we expect other nanodiamonds of the same kind to exhibit an analogous behavior with increasing temperature.
    
	The suggested mechanism for the $\mathrm{NV}^-$ to $\mathrm{NV}^0$ charge-conversion process is tunneling, mediated by substitutional nitrogen atoms in the diamond lattice \cite{Capelli.2022, Manson.2018}. 
	This effect depends on the nitrogen concentration in diamond \cite{Manson.2005, Capelli.2022}, and is not influenced by an increase of the temperature in our sample. 
	The temperature dependence on NV-center charge conversion has previously been investigated for temperatures below room temperature \cite{Bhaumik.2019, Guo.2020}.
	While charge conversion did not show any temperature dependence in \cite{Guo.2020}, an exponential increase of the $[\mathrm{NV}^-]/[\mathrm{NV}^0]$ ratio with decreasing temperature was found in a different sample in \cite{Bhaumik.2019}.
	Since we examine the temperature dependence of charge conversion above room temperature, our findings conform with both observations mentioned above.
	To get an insight into the temperature dependence of charge conversion in our sample below room temperature, measurements at cryogenic temperatures will have to be performed. 

    \begin{figure}[]
            \includegraphics[width=86mm]{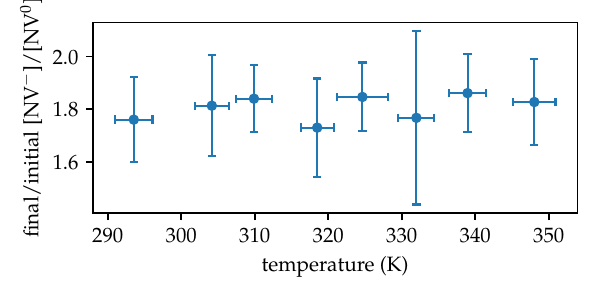}
            \caption{\label{fig:Ratios_T}
            Charge-state ratio of $[\mathrm{NV}^-]$ to $[\mathrm{NV}^0]$ at the longest $\tau$ divided by the ratio at the shortest $\tau$ as a function of the temperature.
            The data is derived from spectral analysis and relaxometry recorded with \SI{0.56}{\milli\watt} laser power.
            }
    \end{figure}
    
    To examine the effect of increasing temperature on charge dynamics, we trace the ratio $[\mathrm{NV}^-]/[\mathrm{NV}^0]$ during the relaxometry measurement at different temperatures.
	For this, we use the results from the spectral analysis to derive $[\mathrm{NV}^-]/[\mathrm{NV}^0]$ during the relaxometry sequence in the signal detection window in the second half of the relaxometry measurement scheme.
	This way, we gain insight into the charge dynamics during the relaxation time $\tau$. Conform with the decrease of the $\mathrm{NV}^0$ fluorescence during $\tau$, we observe an increase of the ratio $[\mathrm{NV}^-]/[\mathrm{NV}^0]$.
	We plot the ratio $[\mathrm{NV}^-]/[\mathrm{NV}^0]$ at the last $\tau$ divided by the ratio $[\mathrm{NV}^-]/[\mathrm{NV}^0]$ at the first $\tau$ for \SI{0.56}{\milli\watt} laser power as a function of the temperature in Fig.~\ref{fig:Ratios_T}.
	We calculate the errors in Fig.~\ref{fig:Ratios_T} from the standard fit errors we obtain from mapping the ratio of $[\mathrm{NV}^-]/[\mathrm{NV}^0]$ to count rate ratios in a fit of form $ax^n+c$ for each temperature.
	The fit is weighted with the errors for the ratio of $[\mathrm{NV}^-]/[\mathrm{NV}^0]$ we obtain from the spectral analysis and the standard deviation of the count rates we measure in the SPCMs.
	As shown in Fig.~\ref{fig:Ratios_T}, the increase of the ratio $[\mathrm{NV}^-]/[\mathrm{NV}^0]$ within the full measurement sequence is around \SI{1.8}{}, and constant with the temperature for a laser power of \SI{0.56}{\milli\watt}.
    Similarly, for a lower laser power of \SI{8}{\micro\watt}, we observe that the increase of the ratio $[\mathrm{NV}^-]/[\mathrm{NV}^0]$ during $\tau$ is independent of the temperature.
	The temperature increase does not affect the charge dynamics during the relaxometry sequence in our sample.
	
	\section{\label{sec:Concl}CONCLUSIONS}
	We conduct temperature-dependent fluorescence spectroscopy and relaxometry of an NV-center ensemble in nanodiamond.
	With this, we examine the temperature dependence of charge conversion between $\mathrm{NV}^-$ and $\mathrm{NV}^0$ in a biologically relevant temperature range from \SI{294}{\kelvin} to \SI{350}{\kelvin}. 
	We perform fluorescence-spectra decomposition at different temperatures and find that $\kappa_{520}$, which describes the photoluminescence ratio between $\mathrm{NV}^-$ and $\mathrm{NV}^0$, is temperature-independent.
	Derived from fluorescence-spectra analysis, we observe that the $\mathrm{NV}^-$ fraction decreases with increasing laser power but is not susceptible to the temperature.
	While we notice a temperature dependence of $1/T_1$ in our sample, the recharge rates $1/T_{R,1}$ and $1/T_{R,2}$ in the dark are not influenced by the temperature increase.
	Therefore, an additional factor in temperature-dependent relaxometry caused by charge conversion between NV-charge states does not need to be considered in applied temperature-dependent nanodiamond relaxometry. 
	Since the substitutional nitrogen concentration influences the charge-conversion process, systematic studies of diamonds of different nitrogen concentrations will give us further insight into the mechanism of this process. 
	Expanding the temperature range from the biologically relevant region to cryogenic temperatures will unravel the temperature dependence in more detail, allowing us to obtain fundamental knowledge about charge conversion.
	
	\begin{acknowledgments}
    We acknowledge support from the nanostructuring center (NSC) of the RPTU Kaiserslautern-Landau.
    This project was funded by the Deutsche Forschungsgemeinschaft
    (DFG, German Research Foundation)—Project-ID No. 454931666.
    Further, I.~C.~B. thanks the Studienstiftung des deutschen Volkes for financial support.
    We thank O. Opaluch and E. Neu-Ruffing for providing the microwave antenna in our experimental setup.
    Furthermore, we thank S. Barbosa for fruitful discussions and experimental support.
    \end{acknowledgments}
	
	\appendix
	
	\section{\label{subsec:decomposition} Spectral analysis}
	Before our spectral analysis, we corrected the fluorescence spectra for wavelength-dependent properties of optical elements in our beam path. Additionally, we subtract a background from each fluorescence spectrum.
	We decompose the fluorescence spectra into basis functions of $\mathrm{NV}^-$ and $\mathrm{NV}^0$ according to Ref.~\cite{Alsid.2019} and follow their nomenclature.
	As described in Ref.~\cite{Alsid.2019}, an area-normalized spectrum $\hat{I}_0^{\mathrm{pre}} (\lambda)$, that we recorded with \SI{4}{\milli\watt} laser power, and an area-normalized spectrum $\hat{I}_-^{\mathrm{pre}} (\lambda)$, that we recorded with \SI{8}{\micro\watt} laser power, are employed to construct the basis function for $\mathrm{NV}^-$ with
	\begin{equation}
	    I_-(\lambda) = \hat{I}_-^{\mathrm{pre}} (\lambda) - \delta_0 \hat{I}_0^{\mathrm{pre}} (\lambda).
	\end{equation}
	
	For unambiguous data evaluation, we optimize $\delta_0$ for all temperatures separately and use their mean as a constant $\delta_0$ for all temperatures. 
	Further, as described in Ref.~\cite{Alsid.2019}, the $\mathrm{NV}^-$ basis function is area-normalized to $\hat{I}_- (\lambda)$ and the basis function for $\mathrm{NV}^0$ calculated with
	\begin{equation}
	    I_0(\lambda) = \hat{I}_0^{\mathrm{pre}} (\lambda) - \delta_- \hat{I}_- (\lambda).
	\end{equation}
	
	The value for $\delta_-$ is kept constant throughout our analysis presented in this paper.
	
	In our previous work in Ref.~\cite{CardosoBarbosa.2023}, we investigated several nanodiamonds of similar composition. We performed the spectral analysis described in Ref.~\cite{Alsid.2019} to obtain $\mathrm{NV}^-$ fractions as a function of the laser power. We verified that our results are fundamentally equal for all examined nanodiamonds. 

    \section{\label{subsec:ZPL}Temperature-dependent spectroscopy}
    
\begin{figure}[h]
            \includegraphics[width=86mm]{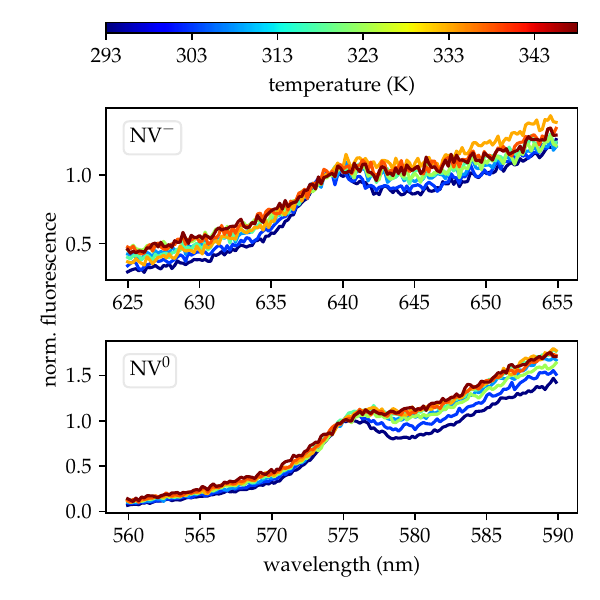}
            \caption{\label{fig:ZPL_T_dep}
            ZPL regions of the decomposed spectra of $\mathrm{NV}^-$ and $\mathrm{NV}^0$, normalized to the spectrum intensity at \SI{639}{\nano\meter} and \SI{575}{\nano\meter}, respectively, for different temperatures. 
            }
    \end{figure}

    For better visualization, we plot the ZPL regions of the decomposed spectra of $\mathrm{NV}^-$ and $\mathrm{NV}^0$ for different temperatures in Fig.~\ref{fig:ZPL_T_dep}. For $\mathrm{NV}^-$ and $\mathrm{NV}^0$, we observe an increase in the ZPL width and a shift to higher wavelengths with increasing temperatures.
    
    \FloatBarrier
		% \bibliography{apssamp}% Produces the bibliography via BibTeX.
		 %%% bibliography
        % \bibliography{Literatur}  
        

	\end{document}